# Evaluating Machine Learning Classifier Approaches, and their Accuracy for the Detection of Cyberattacks on 5G IoT Systems.

**Adem Rosic, rosica@sunypoly.edu**
**State University of New York Polytechnic Institute**
**Utica, NY 13502, UNITED STATES**

***Abstract*—** As 5G continues to expand its coverage and use. Innovative ideas/technologies continue to be implemented within. New vulnerabilities appear, thus resulting in new methods of mitigation and detection to occur. With the architecture that 5G can implement, DDoS (Distributed Denial of Service) is at a higher risk. There are many methods and approaches to help combat this challenge, most of which are implemented in networks containing Wi-Fi (Wireless Fidelity). This article aims to discuss the possible approaches that could be included in 5G technology. The method we will discuss involves Machine Learning. We have used three classifiers to test on datasets (Naïve Bayes, UltraBoost, LogitBoost) with multiple cross-folds, verifying which would have the highest accuracy with multiple factors (such as the cross-folds, verifying whether the number of folds affects accuracy), expanding upon [25] by using feature selection to obtain more accurate results.

# I. Introduction

5G is the fifth-generation, mobile broadband network. It is the successor to 4G LTE (Long term evolution) technology [1]. Its standards are created with careful consideration under the 3GPP, which stands for the 3rd Generation Partnership Program. 3GPP is a standards organization that participated in the development of 5G, and its predecessors like 4G (LTE). 5G was officially produced for the public in 2019. Now, its population of users rapidly increased to an expected population of 1.3 billion active clients by the end of 2022 [2]. One of the many types of infrastructure that will benefit astronomically from 5G networks is the Internet of Things (IoT). IoT is a term that describes “a system of interrelated computing devices, mechanical and digital machines, objects, animals or people that are provided with unique identifiers (UIDs) and the ability to transfer data over a network without requiring human-to-human or human-to-computer interaction.” [3]. Examples of this could be a smart vacuum, that operates on its own, and provides alerts based on events that occur to the machine, whether that is having its trash bin full, or if the device is broken. Another example that is vastly different from a smart vacuum is autonomous cars. The definition “autonomous” defines as “undertaken or carried on without outside control” [4], thus describing autonomous cars as vehicles that can make their own decisions without any outside control, the outside control being the driver, or any controller of that matter. While autonomous cars are not fully established as of now. We know that smart vacuums, and other IoT devices are already in use. Thanks to our existing 802.11 networking, which brings us Wi-Fi, they are in use. With the existing implementation, we can understand IoT and its requirements for future development, concerning cellular networking, like 5G.

One of the issues IoT faces is using outdated software in its systems. As of 2021, there are more than 10 billion active IoT devices [5]. It is said that IoT devices have embedded software that is not patched/hardened. Patching is applying an update on software that handles a vulnerability that is present within the previous version. Hardening includes the process of patching, but also other actions needed to secure the devices' operating system. An example of hardening is the change configuration files for an application, strictly opening ports that are required by the specific application, removing unnecessary services, and compilers, and configuring DAC (Discretionary Access Control), RBAC (Role-Based Access Control), and MAC (Mandatory Access Control). Many IoT devices lack in these areas, widening the Attack Surface (total number of attacks), thus making IoT devices vulnerable and susceptible to a variety of attacks. [6] reports that 900 million, out of the 1 billion cyber-attacks, are said to be related to IoT devices.

Given the large quantity of IoT devices infiltrated, this makes it quite an attractive appearance for hackers whose goal is to create a botnet. A botnet is a malicious network that is operated by an adversary. The client devices are considered “zombies” in these networks. IoT devices are ideal for most hackers to exploit and inject C2 (command and control) software into, turning the device into a zombie in the botnet. Once an adversary creates a botnet its possibilities are endless. A botnet could be used for spyware, proxying, pivoting, and DDoS (Distributed Denial of Service) attacks. [7]

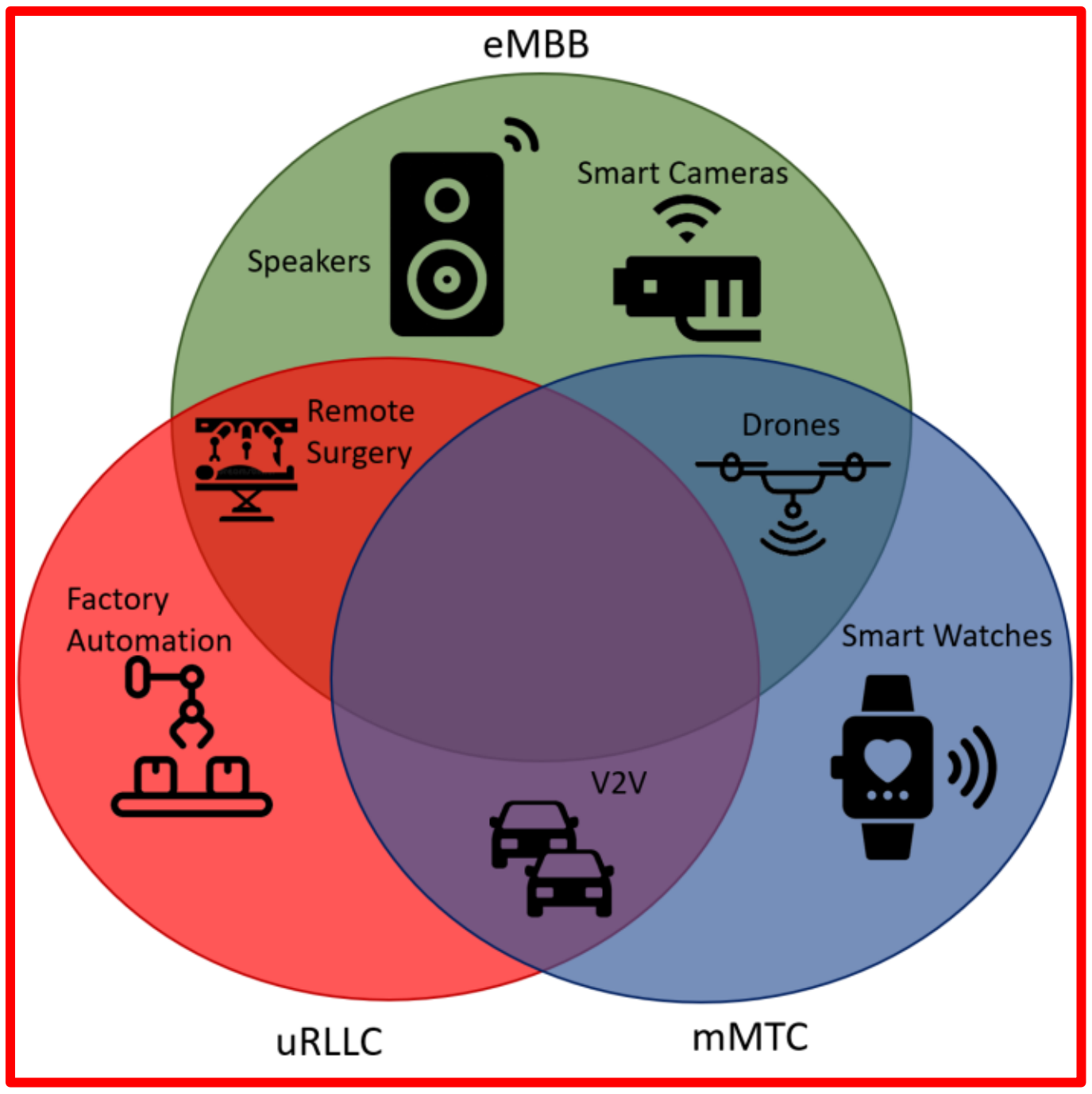

**Figure 1** *Visual IoT use in 5G network and its roles.*

To understand what a distributed denial of service attack is, DoS (Denial of Service) needs to be understood as a prerequisite. Denial of Service attacks are attacks made to deny the action of providing/receiving service. The result of the attack could be temporary or indefinite. There are variations of the DoS attack:

(i) **Yo-yo attacks.** This attack focuses on services that run behind autoscaling, which means as more resources are used, more resources are produced to manage the incoming traffic and resource consumption [8]. This attack aims to disrupt resources momentarily, making this a DoS attack. More importantly, it aims to burden the service provider financially.

(ii) **ARP spoofing.** This is a common attack based on the fundamental concept of the ARP protocol. With the ARP protocol, a client sends a layer 2 broadcast message asking, "what MAC address belongs to this IP?". The MAC address that is related to the IP address, identifies itself with that MAC address. Adversaries can take advantage of this with a race condition. The attacker must respond to the client faster than the actual server. If successful, a Denial of Service occurred, allowing the attacker to provide its service (or nothing at all) as opposed to the desired service the client requested.

(iii) **Wi-Fi de-authentication attacks.** This is a common DoS mechanism that occurs within wireless networks. The objective for the attacker is to send a de-authentication packet to the AP (Access Point) and force the AP to respond to the victim with a de-authentication packet. Disassociating the client from the wireless network.

A Distributed Denial of Service attack enhances the Denial of Service by utilizing multiple devices to attack a particular target. It follows the same fundamentals as a Denial-of-Service attack, however, significantly increases the likelihood of the target's resources being used since attackers are horizontally expanding their resources by using multiple clients, which scales faster than vertically scaling. This brings heavy concern for 5G networks and IoT due to the vulnerability mentioned previously about IoT devices.

Although Cellular networks behave similarly compared to wireless 802.11 networks, the number of clients a cell must handle within a cellular network exceeds the number of clients a regular access point will handle. We will tackle the architecture of 5G's cellular networks, which will include the RAN, Core network, etc. We will also tackle the technologies used that set aside 5G from most networks. Detection of DDoS methods from previous works of literature will be mentioned and demonstrated. Lastly, a discussion of the overall project will occur, discussing the results.

## II. BACKGROUND

As said in the introduction, 5G is the successor to 4G and is the fifth-generation mobile broadband network. Many implementations of 5G are under guidance from the 3GPP, which sets standards and contributes to the development of all generations of cellular networks. 5G is made with three objectives in mind. These objectives are:

(i) **eMBB (Enhanced mobile broadband)**. This term means that higher throughput, capacity, and speeds need to be considered when deploying 5G networks.

(ii) **URLLC (Ultra-Reliable Low Latency)** connections need to be thought of for the sake of mission-critical devices. For example, future hospital machines may need low latency when remote commands are issued. 5G technology aims to provide such a goal.

(iii) **mMTC (Massive Machine-Type Communications)** Is essential in 5G networks for the sake of connectivity density and for machines to communicate with each other efficiently [9]. An example of this is autonomous cars communicating amongst themselves.

5G utilizes many technologies to maintain these goals, such as massive MIMO, beam forming, mm waves, etc. to understand massive MIMO, we must understand MIMO. MIMO, which stands for multiple in multiple out is a technology where multiple antennas and radios are used as a transmitter and receiver, allowing multiple signals to be sent/received from multiple paths, creating less margin for error, thus resulting in higher data rates and range due to the accuracy increasing [10]. Massive MIMO enhances this technology through spatial diversity, which sends that same message through the air, from multiple paths. Interference will not be an issue since multiple copies of the data are sent from multiple directions/paths. Beamforming is the idea where antennas use certain advanced technologies that allow radio messages to be sent in one direction/path as opposed to it be broadcasted through multiple paths [11].

5G aims to achieve the three goals mentioned previously, however not all clients require all three objectives. 5G offers different frequencies to resolve these issues and even more by performing network slicing as well. 3GPP defined the technology for air transmissions in 5G called New Radio (NR). NR is divided into 2 frequency ranges, FR1 (frequency range 1) below the 6 GHz (gigahertz) range, and FR2, which has a range of 24-54GHz. The reason for the different frequency ranges is due to the nature of electromagnetic waves. When the frequency of a wave is higher, the distance the wave can travel is shorter, and objects absorb higher frequencies easier than lower frequencies. FR1 is usually divided into two unofficial bands called the low band, and the high band, representing the low and higher halves of the FR1 range, respectively. With FR2, mm waves are used in higher frequencies and shorter ranges, due to their natural limitations [12].

When devices connect to the cellular network, that usually means they are connected to the RAN of the network. RAN, which stands for Radio Access Network is the part of the cellular network where the cell is speaking to the user equipment and vice versa. RAN can be thought of as the part in a cellular network topology where the air medium begins and takes place [13]. The cells in a RAN are responsible for connecting the user equipment to a core network.

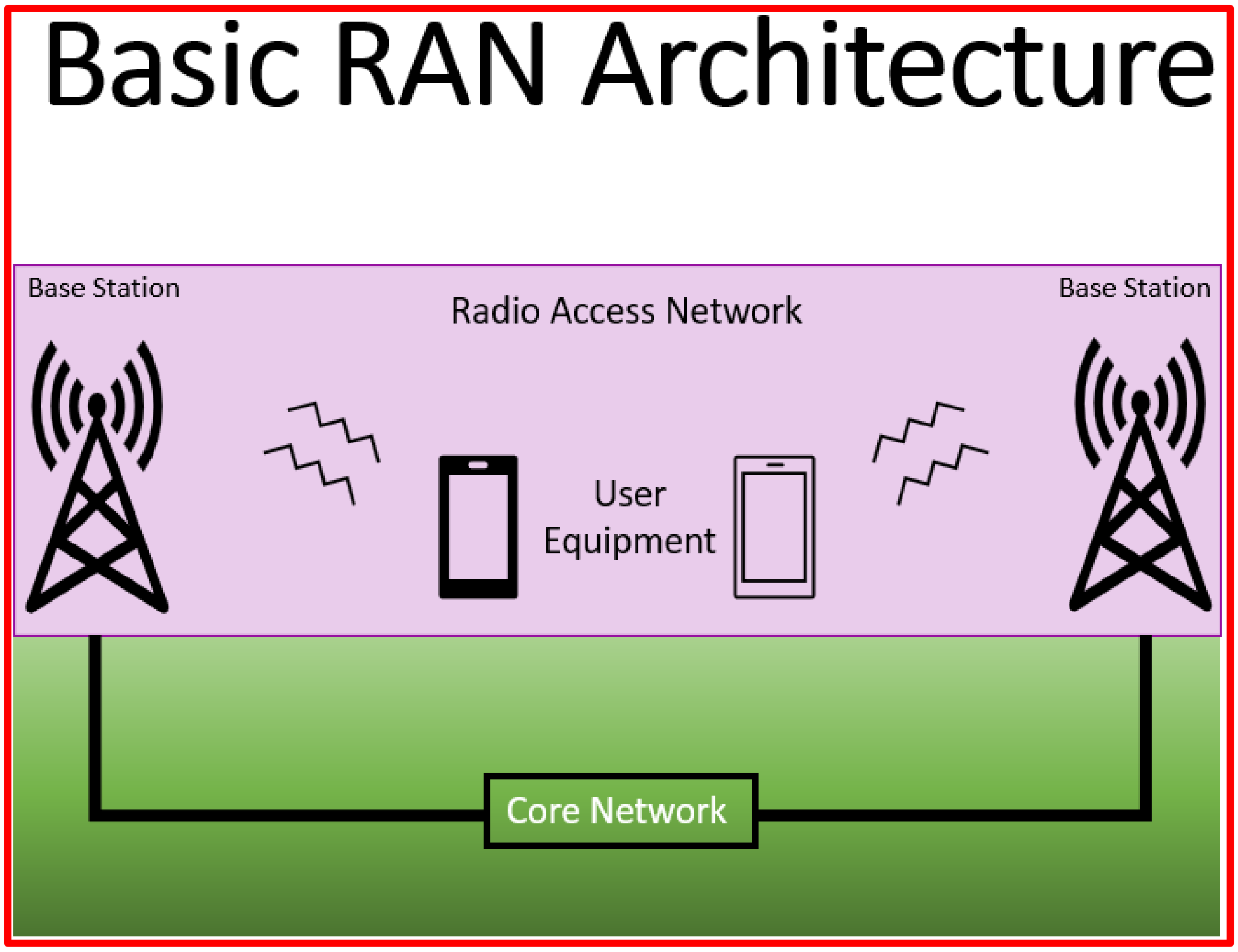

**Figure 2** *Visual of RAN architecture*

After the RAN, we have the core network. The core network is the central network in cellular, that allows subscribers services that should be provided to them [14]. The core network is also responsible for handling requests from the user equipment to external data networks (an example of a data network is the internet). The core network contains services such as:

(i) **AMF (Access and Mobility Management Function)** is the single point of entry for the UE connection to the control plane. Based on what service the UE requests, the AMF sends to UE connection to the **SMF (Session Management Function)** to handle UE requests for services

(ii) **UPF (User Plane Function)** is used to transport the IP data traffic between UE and the data networks

(iii) The **AUSF (Authentication Server Function)** is used to allow authentication and authorization between the UE and services from the 5G core.

(iv) Other functions like **SMF (Session Management Function)**, **PCF (Policy Control Function)**, **AF (Application Function)**, and **UDM (Unified Data Management)**, which are more related to PDU sessions in 5G networks, which are out of scope for this writing [15].

Now that the fundamental concepts of 5G are discussed. We may move on to detection mechanisms that can be used to detect DDoS attacks.

## III. Literature Study

### *a) Feature Selections*

[14] states that many works of literature that discuss DDoS detection methods via Machine Learning are quite popular. However, there seems to be an insufficient quantity of reports that include feature selection. Feature selection is a method where the number of input variables is reduced when developing a model. The general result of this method is less computational time being used to make decisions [15]. This result would be inversely proportional to the result of time to detect malicious traffic. Multiple algorithms were tested on a custom 5G dataset. The stacking algorithm was the most accurate algorithm with all 55 features, showing an accuracy of 97.264%. When feature selection was applied, which decreased the number of features used, [14] noticed that the stacking algorithm would only be 0.083% less accurate. Nonetheless, this has proved that with Machine Learning techniques for detecting DDoS attacks, Feature Selection would be ideal since computational speed is kept at seconds compared to without Feature Selection.

### *b) OMNeT++ IDS against SYN Flood DDoS*

OMNeT++ is a c++ simulation library, with its primary purpose being used for network simulation [16]. [17] utilizes OMNeT++ and creates an IDS out of the framework that is set to detect SYN flood attacks. SYN flood attacks work when the attacker performs two parts of the TCP three-way handshake, however, never completes it, leaving the server's resources busy, and throttling the maximum connection it may be able to handle, leaving innocent clients with no service delivered to them at their request. [17] managed to create an SDN-based IDS that can fit into the routing of the 5G network and was able to perform a SYN Flood attack. The results proved that with 70 UE devices in use of the DDoS attack, the custom IDS was able to maintain accuracy above 90%.

### *c) Detection Accuracy against Steganography in 5G networks*

Steganography is a technique used to hide additional data within the original message. This can be done with mass variations of data (photos, videos, text files, audio, etc.). The purpose of this is to "hide in plain sight" by sending innocent messages, with the malicious message being hidden underneath. Attackers use this to exfiltrate data that is not supposed to be transferred outside of certain networks. This can be used in massive IoT networks, under large botnets, and may use 5G cells/core networks' resources under unnecessary use. [18] used an efficient ANN technique to detect audio files that were modified via steganography with 97.92% accuracy.

### *d) Prediction and Detection of FDIA and DDoS Attacks in 5G enabled IoT*

[19] discusses multiple solutions for securing the 5G architecture for the sake of IoT devices. The security proposals are requested to protect against FDIA (False Data Injection Attack). FDIA is done when an attacker inputs data in a subtle fashion, causing the results to appear different than intended with no error alerting the action [20]. [19] introduces a hierarchical architecture along with a Markov model method, which proves to be a valid solution to prevent DDoS and FDIA. The architecture

included three layers of defense. That being the access layer, the MEC layer, and the cloud layer. The results of this architecture simulation showed a low error rate and a high detection rate of DDoS and FDIA

*e) A wireless Intrusion Detection for the Next Generation (5G) Networks*

The authors, Ferrucci, Richard; Kholidy, Hisham, showed concern for 5G security architecture. This concern was raised with the realization that 5G networks will be handling more data than ever anticipated. The authors used a machine learning classifier, which is an algorithm that orders data into separate classes, or groups. For example, a certain classifier can categorize spam emails to one group, and non-spam to another group [21] LogIT boost classifier, with feature selection used for the reasons stated in the “feature selection” sub-section. The authors simulated D2D (Device2Device) connections and used the AWID dataset to view the accuracy of detection amongst intrusion detection. The results showed that the LogIT classifier had the 2$^{nd}$ highest percentage of precision without feature selection and the worst precision with the CFSsubset and BestFirst feature selection used, but when using the Ranker Feature Selection, the precision was again the 2$^{nd}$ highest for LogIT. The purpose of the literature was to prove what was discussed in the “feature selection” sub-heading. Which was that feature selection produces results of higher quality while minimizing the time of training [22].

*f) 5G Networks Security: Attack Detection Using the J48 and the Random Forest Tree Classifiers*

The authors of this literature, which are Bruce Steel II and Dr. Hisham A. Kholidy, experimented with the J48 Decision Tree classifier, and the Random Forest Tree Classifiers. The difference between a decision Tree classifier vs a Random Forest Tree classifier is that a decision tree is built from the entire dataset. Whilst a random forest tree randomly selects specific rows and variables and forms multiple decision trees based on values recognized. Once the multiple trees are made, while observing, the class that has the most votes out of the random forest, is chosen, putting it in that class [23]. The authors used J48 as the decision tree classifier, comparing the results to the random forest classifier used. The dataset used was not AWID, as used in the previous sections. Instead, the UNSW-NB15 dataset [38,39] was used. The similarity between most datasets is that the datasets were made for regular 802.11 wireless network traffic. This choice is the most feasible option since 1. No publicly available dataset catered to 5G networks, and 2. 802.11 networks share similarities with cellular networks. The WEKA software was used, which is an open-source, machine learning software that is made in java [24]. After experimenting, it was revealed that accuracy stood at 85%. Compared to other works of literature stated previously, this happened to show the least accurate results. Nonetheless, the fact that 85% of objects were classified correctly sets a high bar regardless [25].

*g) DDoS detection in 5G-enabled IOT networks using Deep Kalman Backpropagation Neural Network - International Journal of Machine Learning and Cybernetics*

The author of this article, Almiani, M., AbuGhazleh, A., Jararweh, Y., and Razaque, A. used the Kalman Backpropagation neural network to determine its accuracy against the CICDDoS2019 dataset. Neural networks, which could be considered artificial neural networks, or simulated neural networks, are a subset of machine learning. There are multiple layers within a neural network, those layers are the node layers, input layers, hidden layers, and an output layer. Data is passed on through each layer, enabling the computer to continuously learn and provide more accurate results over time [26]. MATLAB 2019b was used when coding the algorithm. The experiments were performed on a MacBook pro. The accuracy was shown with the highest of 96.02% in one of the sub-datasets, and the lowest was 90.32% in another one of the sub-datasets. The purpose that this work of literature served was the fact that neural networking is a qualified candidate to implement in 5G networks to detect malicious data being sent [27].

*h) Secure5G: A Deep Learning Framework Towards a Secure Network Slicing in 5G and Beyond*

The authors of this work of literature, which go by the names of Anurag Thantharate, Rahul Paropkari, Vijay Walunj, Cory Beard, and Poonam Kankariya, experimented with methods to harden 5G network slicing. Network slicing is a method that serves the purpose of creating multiple logical networks within an infrastructure. As said before, many clients in a 5G network may need one of the three goals that 5G strives to supply (URLLC, eMBB, mMTC). We can limit the capabilities of those clients, to support a specific use case. For example, hospital equipment may not need eMBB, however, URLLC and mMTC are essential to them. A network slice, catered to that need, may be created for the User Equipment [28]. The authors developed a Neural Network version of Network Slicing, to eliminate potential malicious incoming connections. The neural network goes by the name of “Secure5G”, which is an extension of the DeepSlice research done by the same authors. DeepSlice is a neural network, which contains both the input layer, the output layer, and 2 hidden layers. The purpose of DeepSlice is the view multiple characteristics of the devices connecting to the 5G network and determining which network slice the user equipment belongs to. The issues that concerned the artists the most are:

(i) **Volume based Attacks**. This attack works like the DoS attacks, exhausting the victims' resources by sending many network packets/connection requests.

(ii) **Masking Botnets (Spoofing) Attack**. This attack functions when an attacker modifies the source of the attack, making the traffic appear legitimate. This attack can be used when a botnet is in place. For example, a mobile phone requesting a network slice that has the capabilities of mMTC, when the actual network slice the smartphone should request is eMBB. This can have adverse effects on clients in the network slice that need the mMTC, the malicious clients can request a large amount of connection, or issue many packets, exhausting the resources in the 5G network, and may hinder the mMTC goal of the 5G network promises. 3GPP has a standard that indicates that clients can assign themselves to multiple slices, which creates another issue since malicious clients can

The authors of this title performed the following attacks under the improved version of DeepSlice, which is Secure5G. Using the DeepSlice dataset, the results show 98% accuracy. This shows that DeepSlice ensures end-to-end security between the 5G network and the user equipment. Several improvements that the authors are actively working on include [29].

(i) Secure UE capability
(ii) SIM Security in 5G
(iii) Unauthorized usage of shared resources between slices
(iv) Network Slice exhaustion

*i) Machine Learning based Anomaly Detection for 5G networks*

Written by Jordan Lam, Macquarie University, Robert Abbas. This work of literature focuses on the proposal of SDS, which stands for Software-defined security. Software-defined security is a term in which an IT environment lacks hardware dependency for security. Each device implemented within the IT environment is automatically adjusted under the security policy created. The result that the authors obtained was 96.4%. [30]

*j) Performance Analysis of Machine Learning Algorithms in Intrusion Detection System: A Review*

Written by T.Saranya, S.Sridevi, C.Deisy, Tran Duc Chung, and M.K.A.Ahamed Khan. This work of literature used the KDD-CUO dataset, with multiple classifiers such as SVM, Naïve Bayes, Decision Table, Decision Tree, ANN, etc. The authors had a training set, and the testing set, and it was used based on the 80-20 rule. The highest out of all the classifiers in terms of accuracy turned out to be the Random Forest classifier, while the lowest being the PCA-LDA-SVM classifier, with an accuracy of 92.16%. Nonetheless, the fact that all classifiers had an accuracy of above 90% shows the future of IDS/IPS. [37]

## IV. Practical Tutorial

Weka was used to experiment on comparing accuracy between multiple Machine learning classifiers. The Bot-IoT dataset [31,32,33,34,35] was used in our practical tutorial. The dataset contains packets that were sent/received from IoT devices, whether they are malicious or not. As said before, IoT devices will be most devices used in a 5G network. The tests were run on an AMD 5600x CPU, NVIDIA GEFORCE RTX 3080, 32GB DDR4 RAM, 1TB SSD, and 2TB HDD. The dataset contains multiple file formats of the dataset, such as pcap, CSV, BRO, and Argus. Weka can read current datasets and display the associated attributes. Once access to Weka has been granted, the next requirement is to enter the "Explorer" application within WEKA.

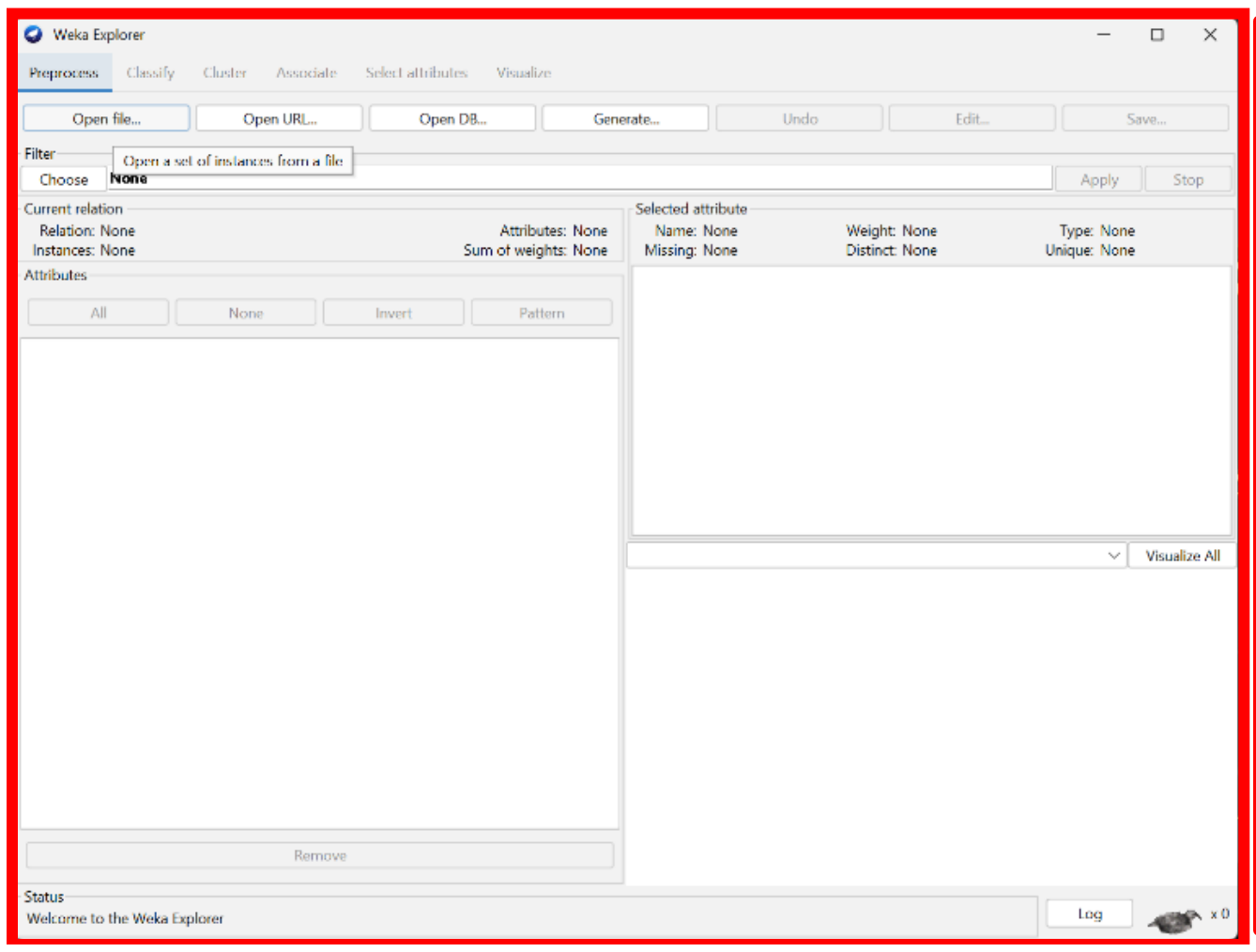

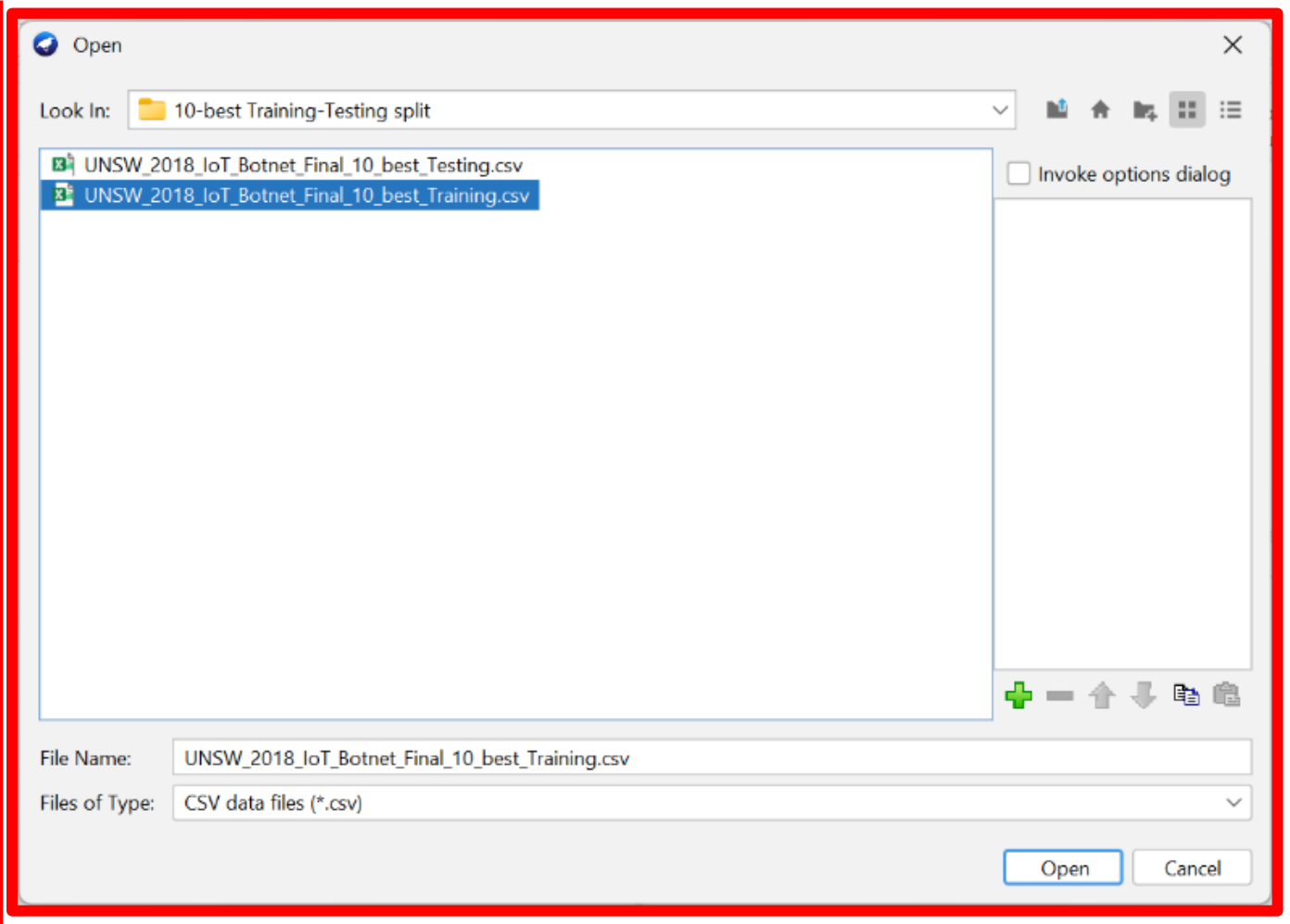

**Figures 3** and **4**, *respectively, show the explorer application in WEKA and a method for selecting a file to run.*

Once entered into the "Explorer" application, we must enter the designated file to input for WEKA to read. The issue with this dataset is the missing/incorrect features inserted in the CSV files provided. When given the dataset, the attribute names were not apparent in the 4 CSV files containing the entire database. There were attribute names given in the training and testing, however, these names were incorrect and did not match the length of the dataset values. Another CSV file is present, explaining all the features possible. The feature list was constructed.

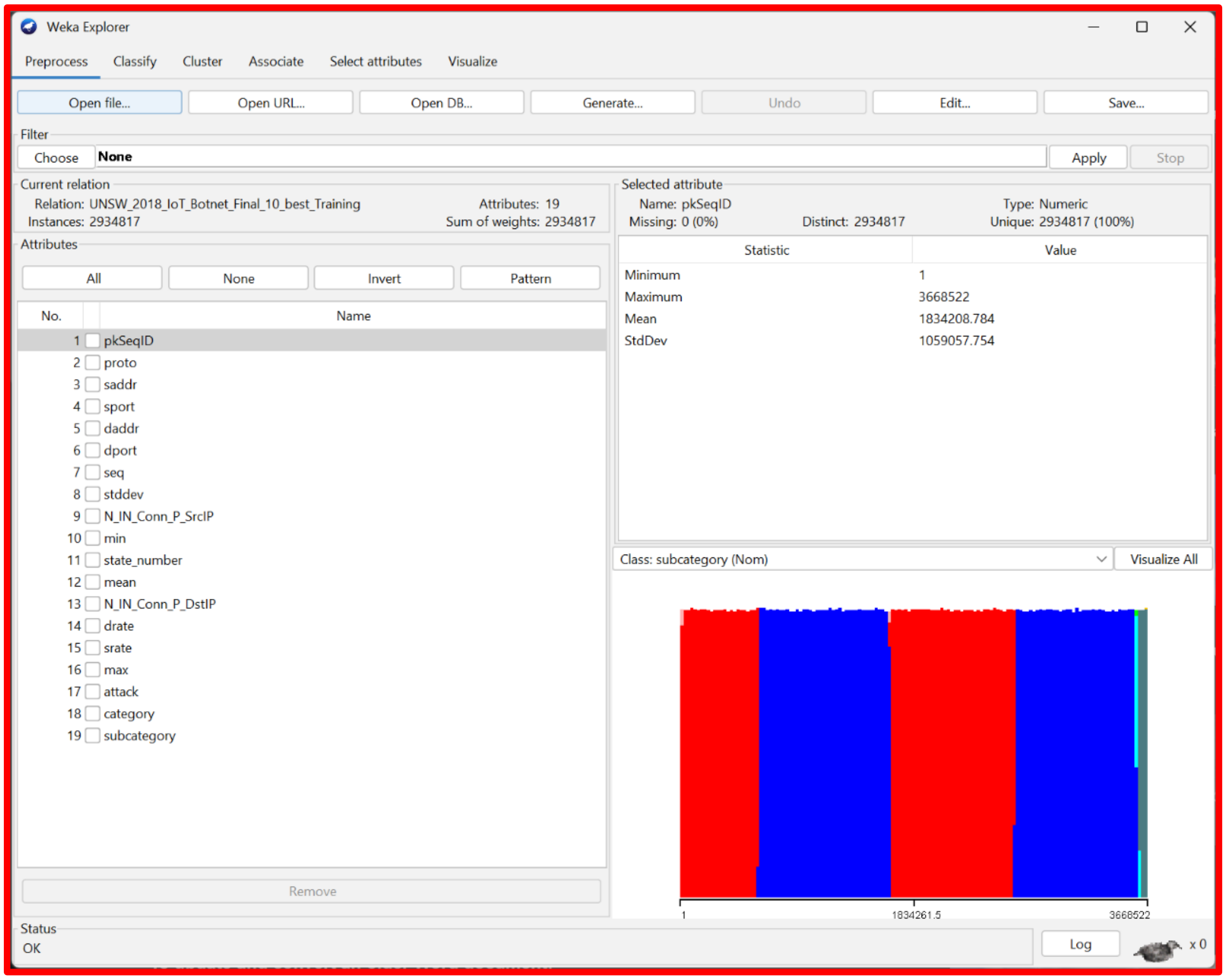

**Figure 5** *WEKA Explorer window, containing the features in the dataset.*

Once WEKA successfully imports the file input, the contents will appear towards the page's bottom-left as seen in figure 5. These are the features that the classifier algorithm observes to justify what label is associated with the subject. Some features are seen to not be required. In this case, it was deemed that the pkseqID, stddev, mean, state_number, attack, and subcategory features were not required. This is because some of the labels deemed an "unrealistic" indicator of a malicious packet, would not be seen in production. To remove these features, click on the checkbox associated with the features, then click the "remove" button. The features will no longer be present. The technique that was just performed is known as feature selection, as discussed in the "Literature Study" section. The "label" feature indicated whether the packet was malicious or not, which may be the reason previous research with this dataset contains accuracy within the 85% and higher range. All other features not mentioned were used as input in the feature selection process for all classifiers.

There are features in this dataset that are considered strings. This interferes with some classifiers, as they do not support string attributes. To convert our attributes to Nominal, which puts our values into "categories," we must select the features in question. In this case, the features were sport and dport. Once selected, click on the "Choose" button that is in the "filter" section of the Explorer application. A filter called "StringToNominal" will be present. Select the desired filter. After selection, click on the filter in the text field, and options will appear as they do in Figure 6. The values in Figure 6 must match.

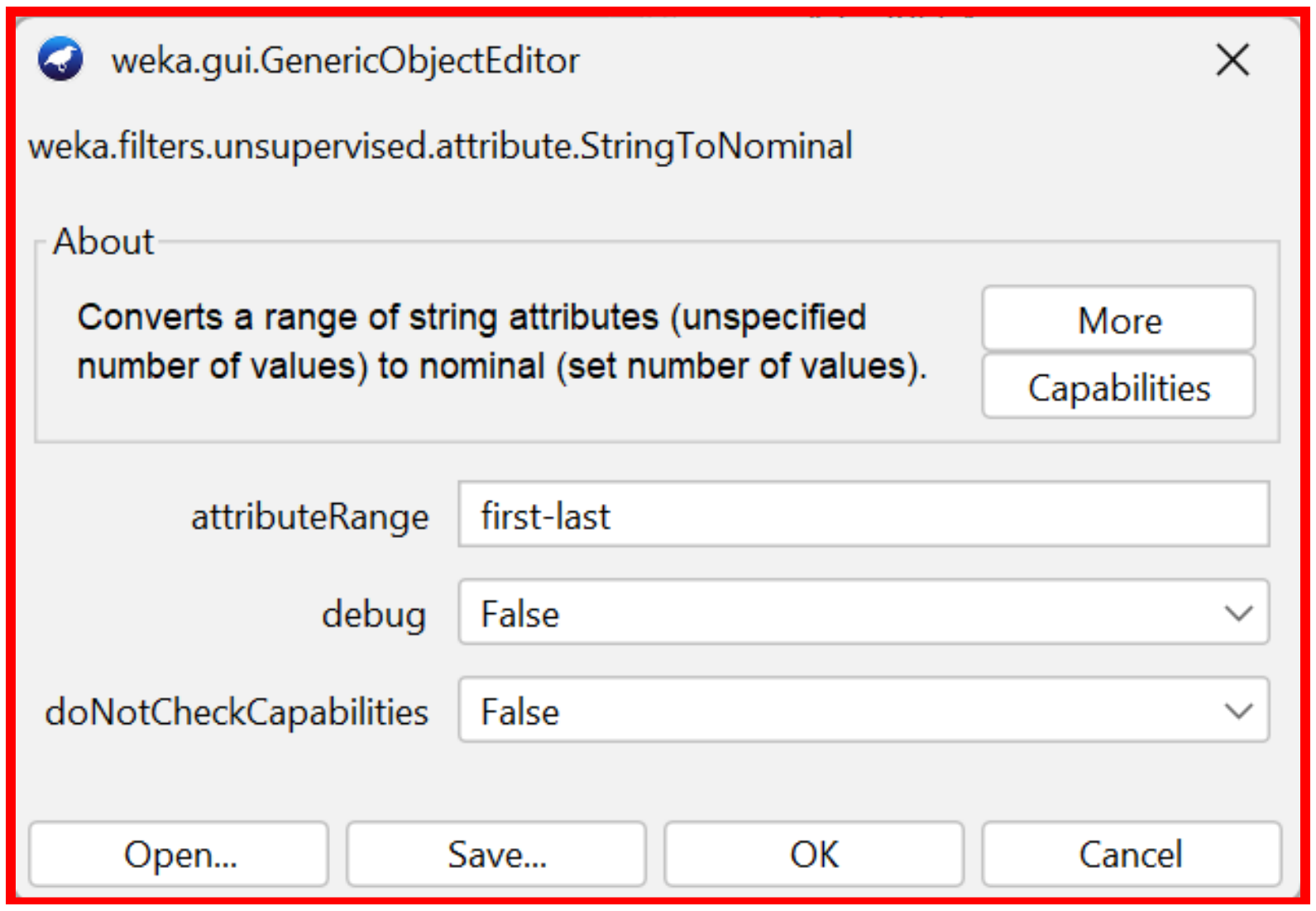

**Figure** 6 *WEKA filter customization values.*

Once the values are entered, click the OK button. The purpose is to set the arguments of this filter to iterate from the first string attribute to the last. Click the apply button. The string has converted to nominals, enabling the capability to use specific classification algorithms. Refer to figure 7 for applying the filter.

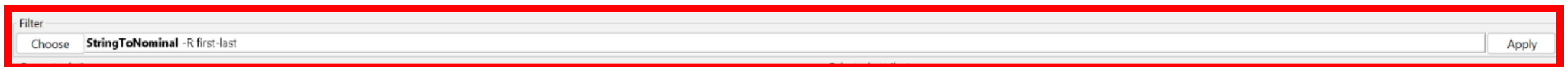

**Figure 7** *WEKA Explorer filter section to apply filter.*

Navigate towards the “Classify” tab in the Explorer application. This tab is the area in which classification algorithms could initiate. The tab displays test options, and multiple classifier models (refer to figure 8 for a visual of the tab). The test options are:

(i) Use training set: This uses the model to test the dataset provided to the application against the model.

(ii) Supplied test set: Uses another test set that must be provided to the application, to use against the model built.

(iii) Cross-validation: This breaks the dataset into ***n*** sections (***n*** is represented via the value of the “folds” field). One of the ***n*** sections will be used as a test set, while the others will be used to build a model to test the remaining set. A loop occurs ***n*** times until all iterations are complete.

(iv) Percentage splits: Splits the dataset by a percentage. One portion is used as the model, while the other is used as the test set.

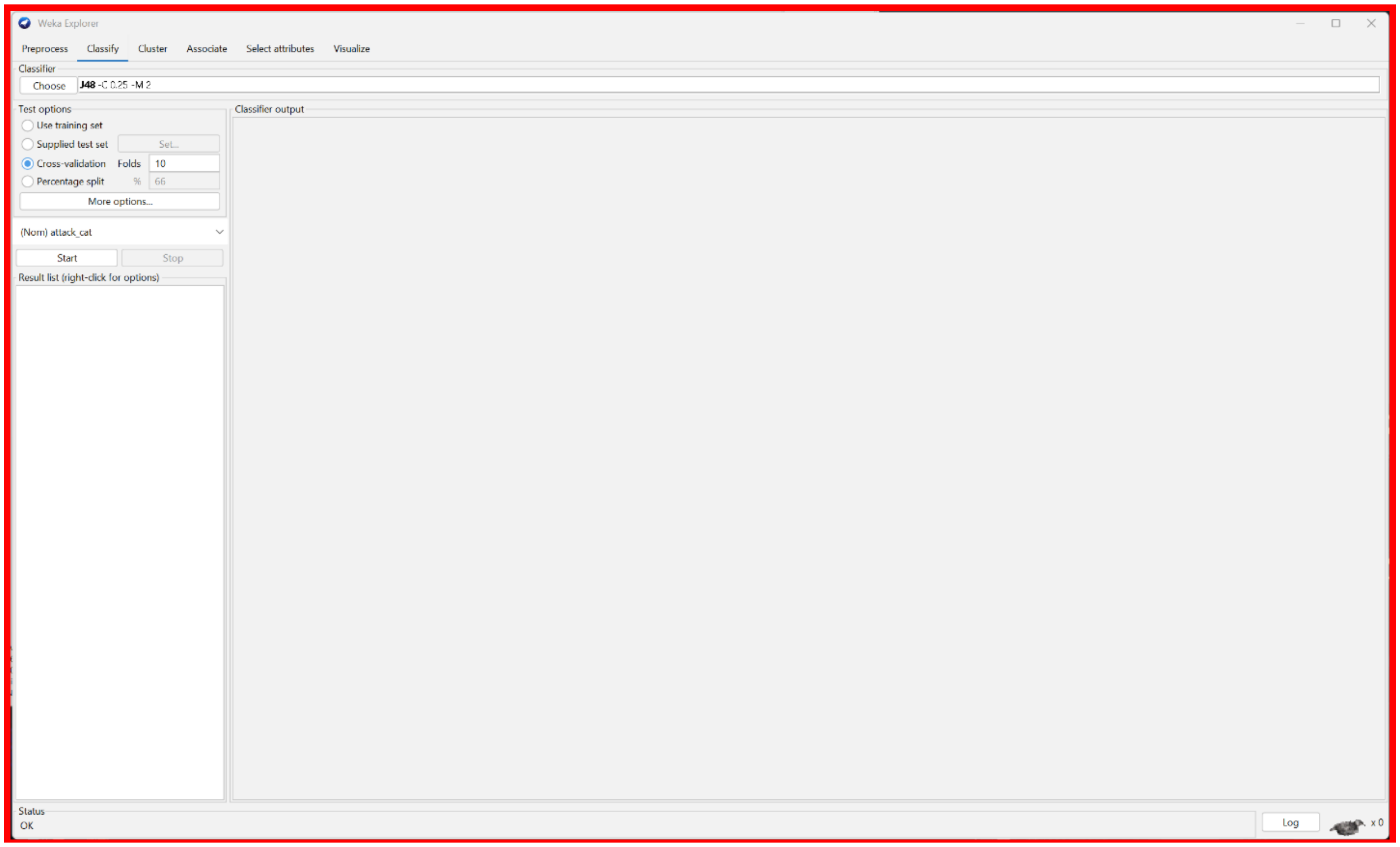

**Figure 8** *WEKA Classify section in the Explorer window*.

As said previously, there are multiple classifiers available. The classifiers that were tested were Naïve Bayes, UltraBoost, and LogITboost. All algorithms were tested via Cross-Validation. To run the tests, click the start button.

## V. Discussion

The experiment is beneficial to those learning the basics of Machine Learning and 5G network fundamentals. This shows the future of cybersecurity technologies. Some difficulties that came across were the datasets, there were multiple datasets that did not have the proper formats that WEKA supported. There were also datasets that contained a quantity that did not produce confident results (e.g., 99% accuracy with only 10k instances). With a higher quantity, a more accurate result can be provided. Another issue is the labels, some labels could be seen as unrealistic. For example, there are labels that go under the name “malicious”, indicating if the packet is malicious or not. This affects the integrity of the result since there are no IP packets in a real-life scenario that labels themselves as malicious. Another issue is computer power. Datasets may be too big for computing power, depending on the user’s spec, and the algorithm in use. For this experiment, the optimal algorithms used were Naïve Bayes, UltraBoost, and LogITboost. These algorithms yielded confident results with the dataset used and the feature selected.

| TP Rate | FP Rate | Precision | Recall | F-Measure | MCC | ROC Area | PRC Area | Class |
|---|---|---|---|---|---|---|---|---|
| 0.998 | 0.142 | 0.886 | 0.998 | 0.939 | 0.870 | 0.958 | 0.934 | DDoS |
| 0.851 | 0.004 | 0.995 | 0.851 | 0.917 | 0.867 | 0.957 | 0.965 | DoS |
| 0.938 | 0 | 0.985 | 0.938 | 0.961 | 0.960 | 0.998 | 0.985 | Reconnaissance |
| 0.922 | 0 | 0.733 | 0.922 | 0.817 | 0.822 | 1.000 | 0.812 | Normal |
| 0.969 | 0 | 0.323 | 0.969 | 0.485 | 0.560 | 1.000 | 0.795 | Theft |

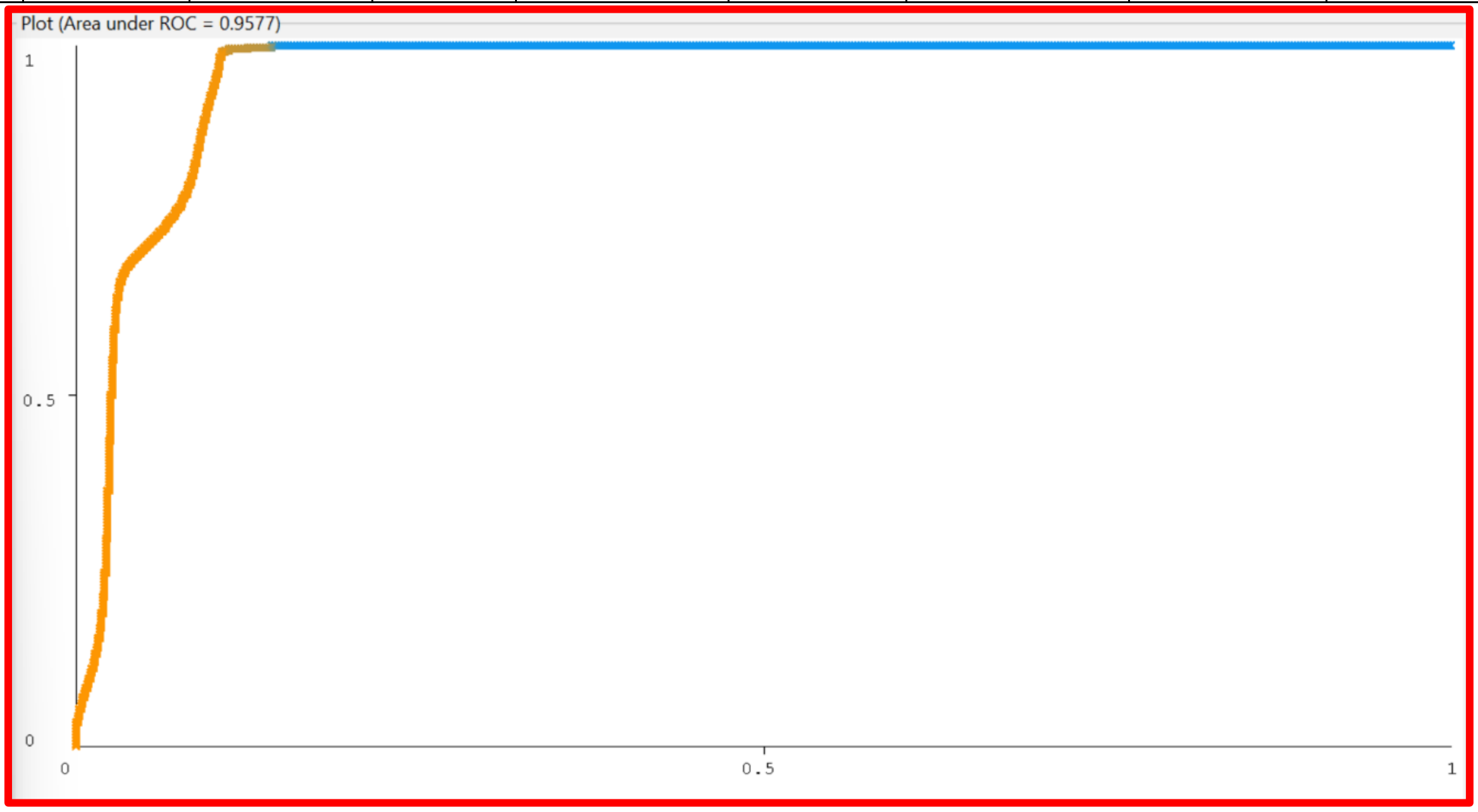

**Figures 9** and **10** *Twenty-fold Naïve Bayes classifier detailed accuracy report per exploit.*

| TP Rate | FP Rate | Precision | Recall | F-Measure | MCC | ROC Area | PRC Area | Class |
|---|---|---|---|---|---|---|---|---|
| 0.998 | 0.141 | 0.887 | 0.998 | 0.939 | 0.870 | 0.958 | 0.934 | DDoS |
| 0.852 | 0.004 | 0.995 | 0.852 | 0.918 | 0.867 | 0.957 | 0.965 | DoS |
| 0.937 | 0 | 0.985 | 0.937 | 0.960 | 0.960 | 0.998 | 0.984 | Reconnaissance |
| 0.919 | 0 | 0.741 | 0.919 | 0.820 | 0.825 | 1.000 | 0.809 | Normal |
| 0.969 | 0 | 0.323 | 0.969 | 0.485 | 0.560 | 1.000 | 0.796 | Theft |

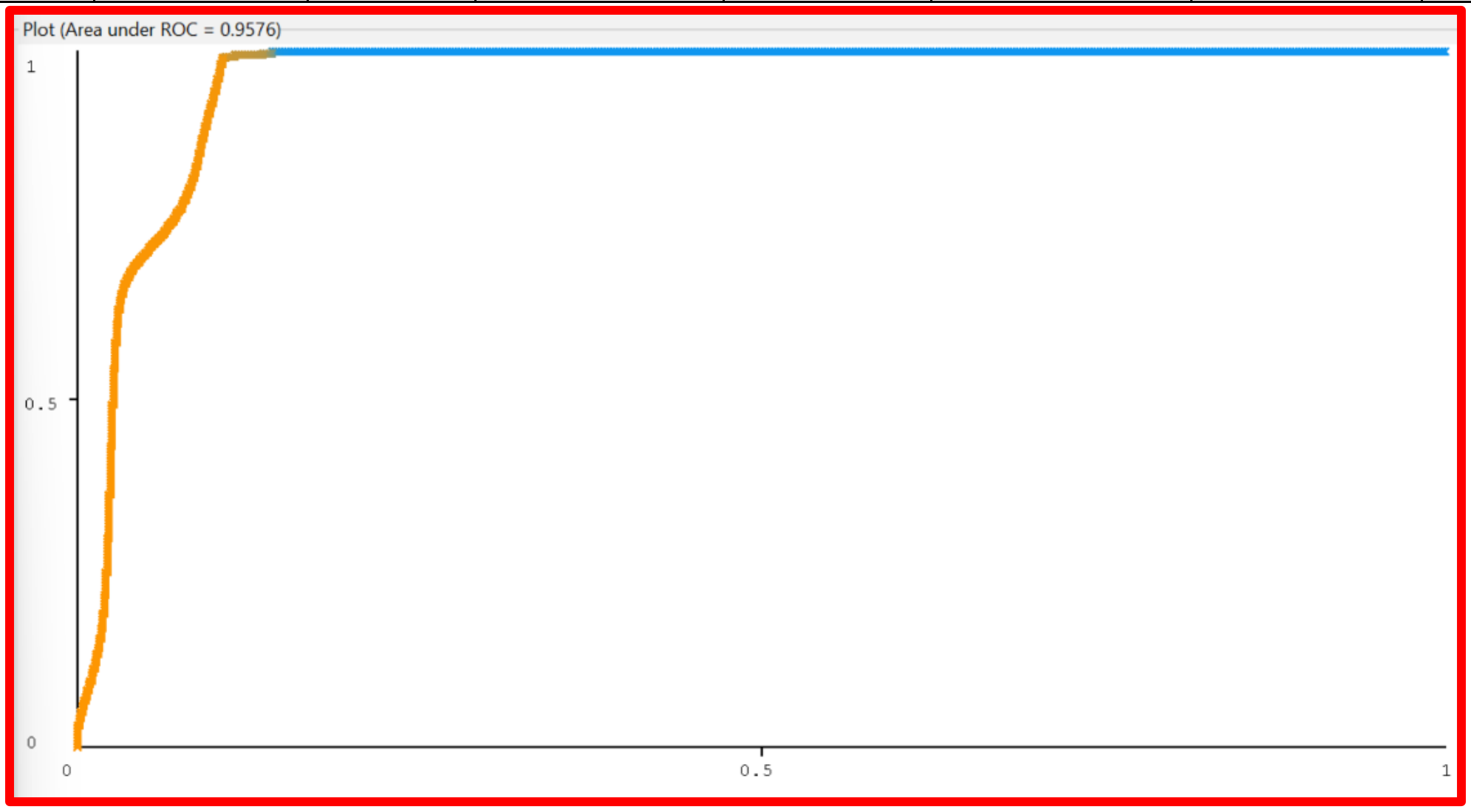

**Figures 11** and **12** *Ten-fold Naïve Bayes classifier detailed accuracy report per exploit.*

| TP Rate | FP Rate | Precision | Recall | F-Measure | MCC | ROC Area | PRC Area | Class |
|---|---|---|---|---|---|---|---|---|
| 0.998 | 0.141 | 0.887 | 0.998 | 0.939 | 0.870 | 0.957 | 0.933 | DDoS |
| 0.852 | 0.004 | 0.995 | 0.852 | 0.918 | 0.867 | 0.957 | 0.965 | DoS |
| 0.935 | 0 | 0.984 | 0.935 | 0.959 | 0.958 | 0.998 | 0.983 | Reconnaissance |
| 0.922 | 0 | 0.751 | 0.922 | 0.832 | 0.832 | 1.000 | 0.794 | Normal |
| 0.923 | 0 | 0.314 | 0.923 | 0.469 | 0.538 | 1.000 | 0.640 | Theft |

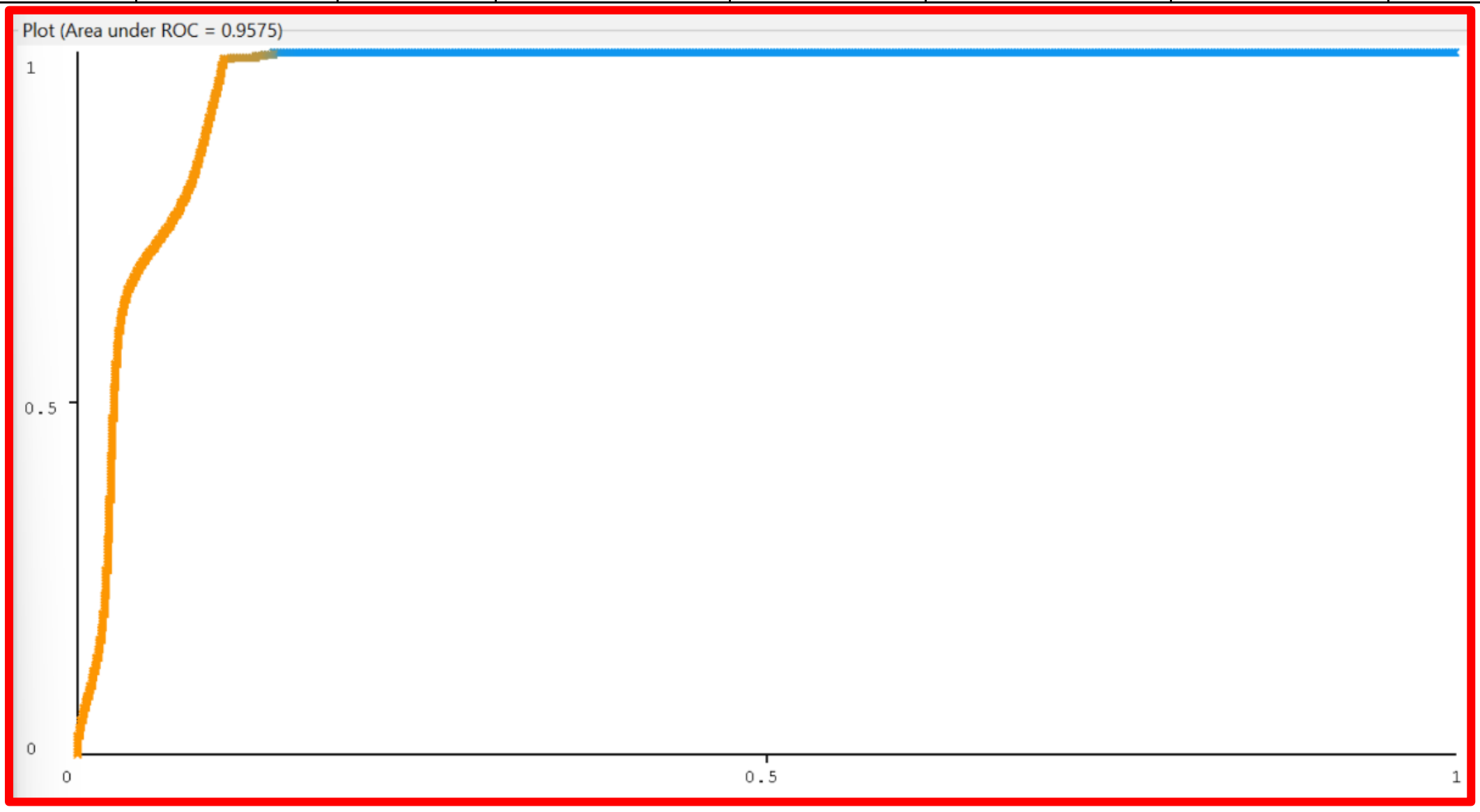

**Figures 13** and **14** *Five-fold Naïve Bayes classifier detailed accuracy report per exploit.*

| TP Rate | FP Rate | Precision | Recall | F-Measure | MCC | ROC Area | PRC Area | Class |
|---|---|---|---|---|---|---|---|---|
| 0.969 | 0.137 | 0.886 | 0.969 | 0.926 | 0.840 | 0.971 | 0.972 | DDoS |
| 0.783 | 0.004 | 0.993 | 0.783 | 0.876 | 0.810 | 0.952 | 0.954 | DoS |
| 0.858 | 0.033 | 0.402 | 0.858 | 0.547 | 0.573 | 0.993 | 0.847 | Reconnaissance |
| 0.838 | 0.014 | 0.008 | 0.838 | 0.015 | 0.079 | 0.994 | 0.116 | Normal |
| 0.985 | 0.004 | 0.005 | 0.985 | 0.011 | 0.072 | 0.999 | 0.040 | Theft |

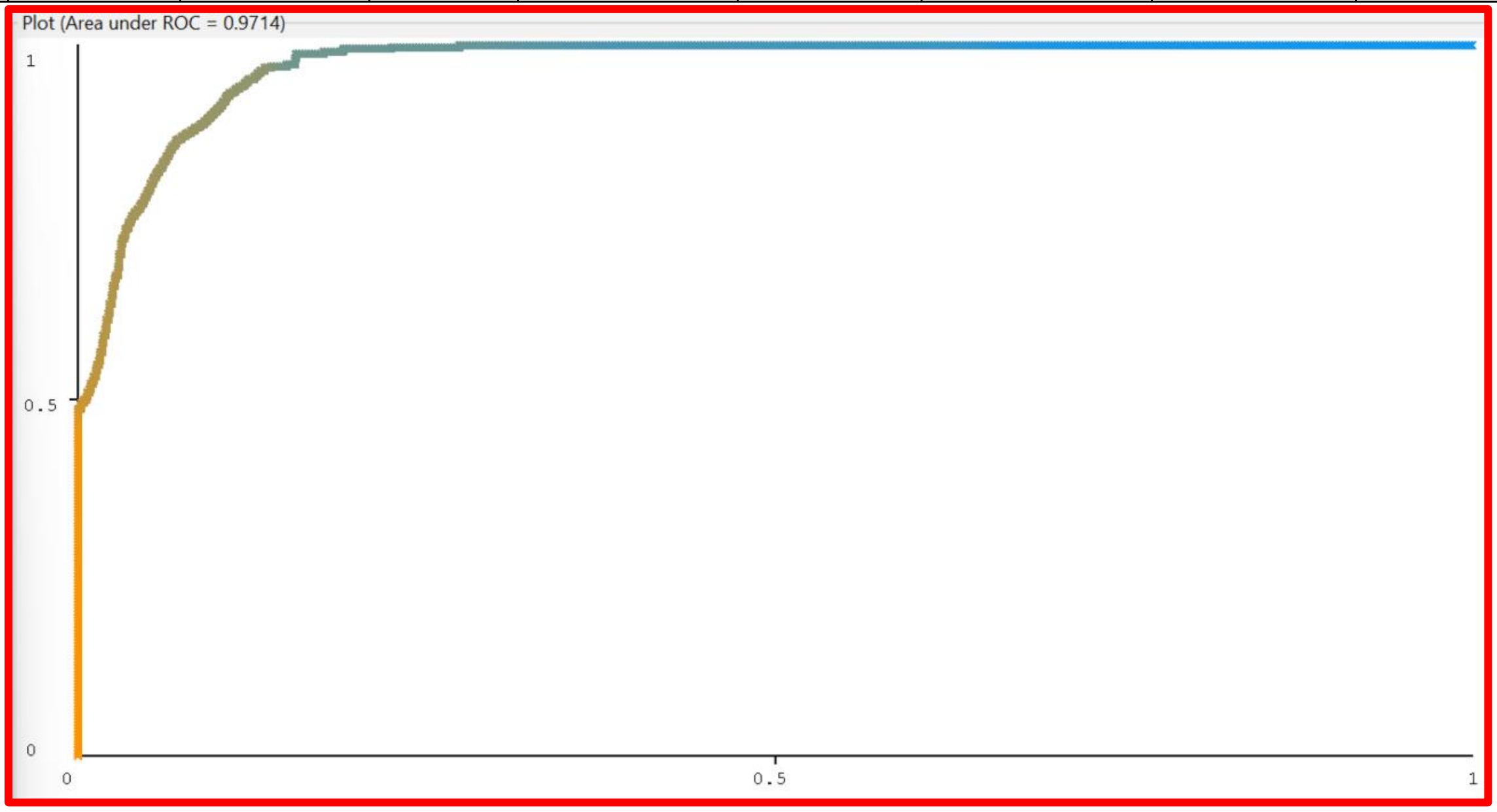

**Figures 15** and **16** *Twenty-fold UltraBoost classifier detailed accuracy report per exploit.*

| TP Rate | FP Rate | Precision | Recall | F-Measure | MCC | ROC Area | PRC Area | Class |
|---|---|---|---|---|---|---|---|---|
| 0.971 | 0.137 | 0.887 | 0.971 | 0.927 | 0.842 | 0.972 | 0.972 | DDoS |
| 0.785 | 0.004 | 0.994 | 0.785 | 0.877 | 0.811 | 0.953 | 0.955 | DoS |
| 0.858 | 0.032 | 0.402 | 0.858 | 0.548 | 0.574 | 0.992 | 0.840 | Reconnaissance |
| 0.830 | 0.012 | 0.008 | 0.830 | 0.017 | 0.083 | 0.994 | 0.113 | Normal |
| 0.985 | 0.004 | 0.005 | 0.985 | 0.011 | 0.073 | 0.999 | 0.040 | Theft |

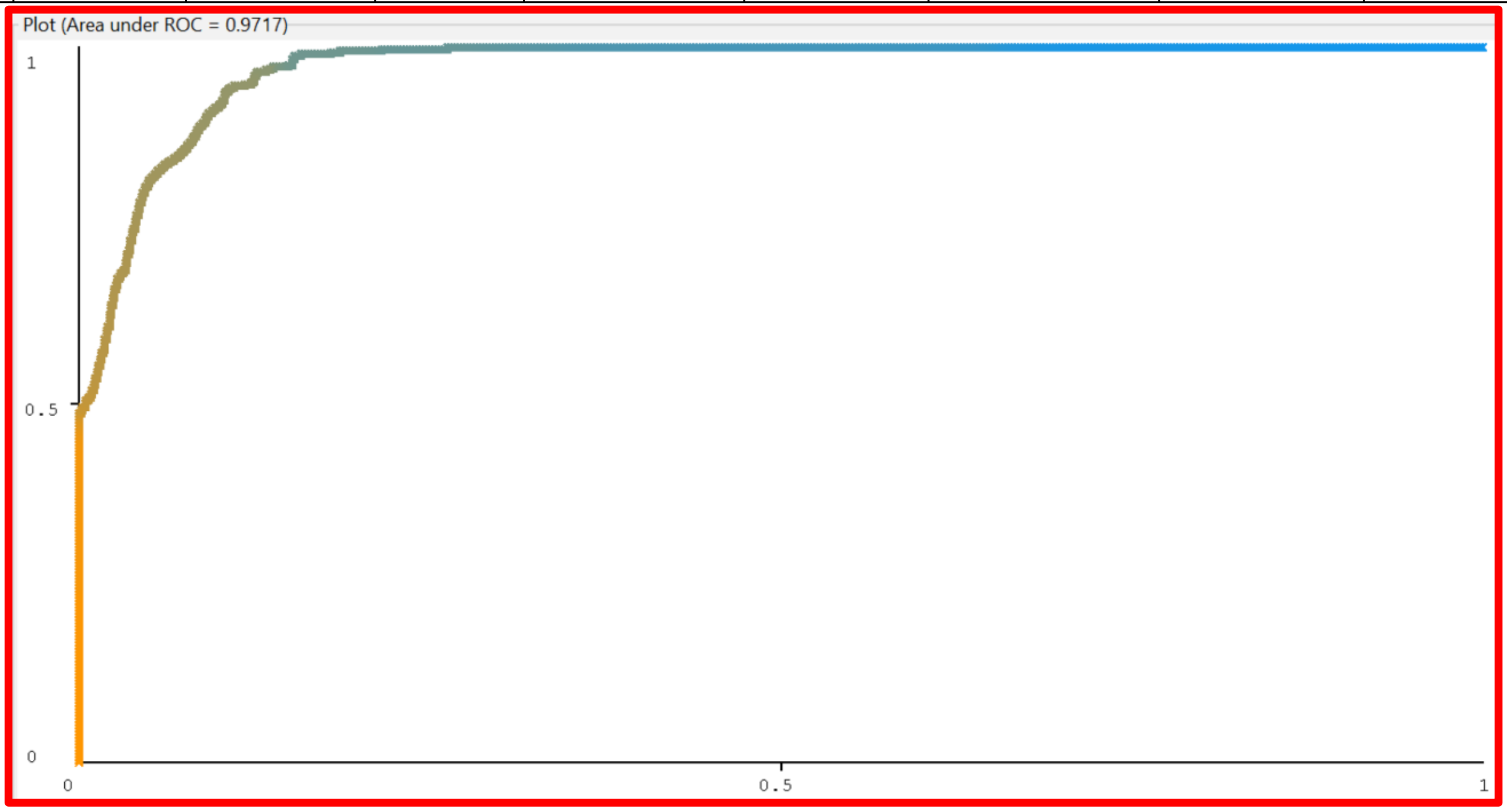

**Figures 17** and **18** *Ten-Fold UltraBoost classifier detailed accuracy report per exploit.*

| TP Rate | FP Rate | Precision | Recall | F-Measure | MCC | ROC Area | PRC Area | Class |
|---|---|---|---|---|---|---|---|---|
| 0.956 | 0.128 | 0.892 | 0.956 | 0.923 | 0.833 | 0.971 | 0.972 | DDoS |
| 0.793 | 0.017 | 0.974 | 0.793 | 0.874 | 0.801 | 0.952 | 0.954 | DoS |
| 0.855 | 0.033 | 0.397 | 0.855 | 0.542 | 0.569 | 0.992 | 0.827 | Reconnaissance |
| 0.824 | 0.013 | 0.008 | 0.824 | 0.016 | 0.080 | 0.994 | 0.113 | Normal |
| 0.985 | 0.004 | 0.006 | 0.985 | 0.011 | 0.074 | 0.999 | 0.037 | Theft |

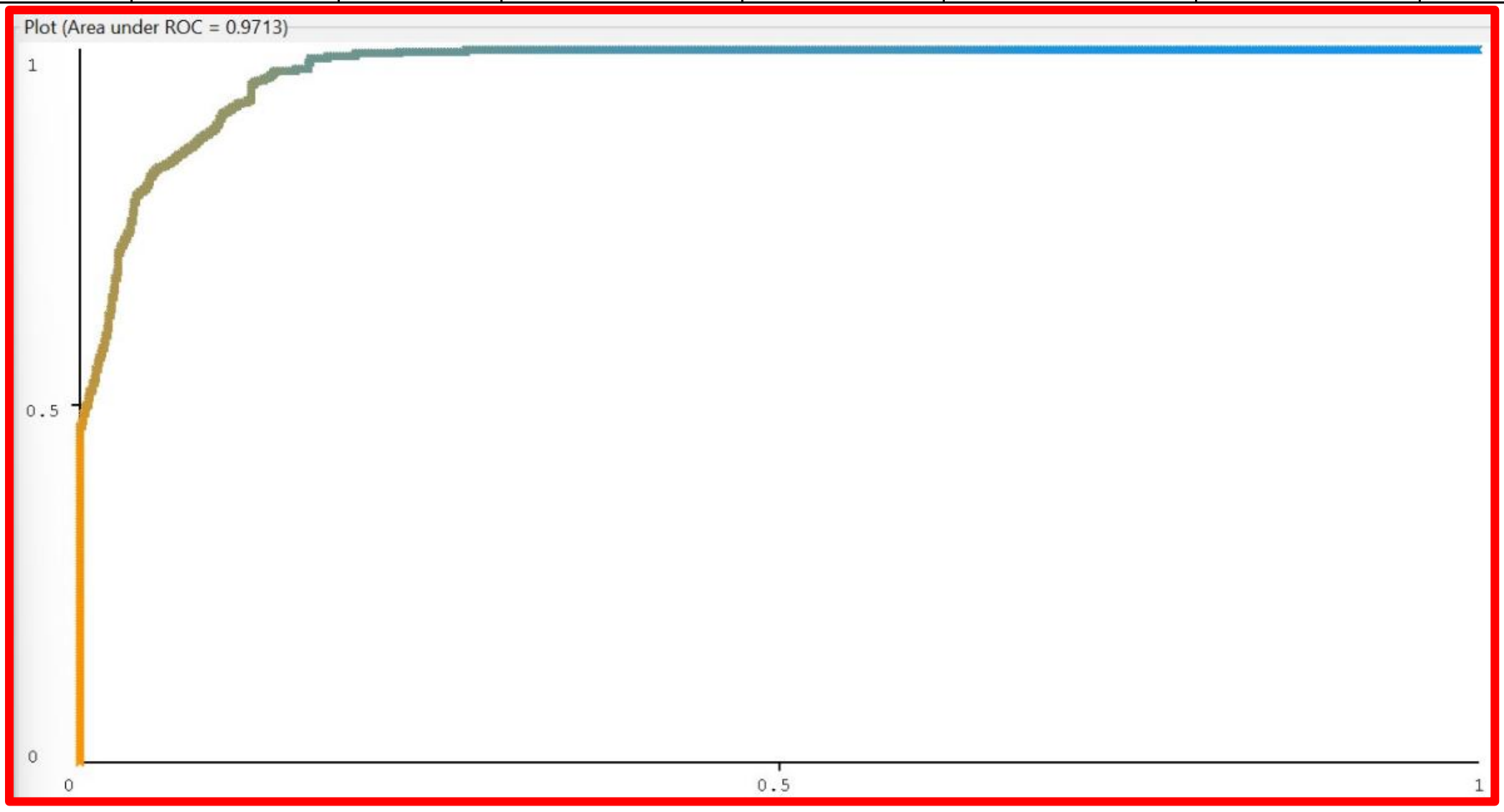

**Figures 19** and **20** *Five-fold UltraBoost classifier detailed accuracy report per exploit.*

| TP Rate | FP Rate | Precision | Recall | F-Measure | MCC | ROC Area | PRC Area | Class |
|---|---|---|---|---|---|---|---|---|
| 0.985 | 0.001 | 0.999 | 0.985 | 0.992 | 0.983 | 0.998 | 0.999 | DDoS |
| 0.999 | 0.015 | 0.982 | 0.999 | 0.991 | 0.983 | 0.999 | 0.998 | DoS |
| 0.992 | 0 | 0.996 | 0.992 | 0.994 | 0.994 | 1.000 | 0.999 | Reconnaissance |
| 0.514 | 0 | 0.856 | 0.514 | 0.642 | 0.663 | 0.993 | 0.693 | Normal |
| 0.846 | 0 | 1.000 | 0.846 | 0.917 | 0.920 | 1.000 | 0.943 | Theft |

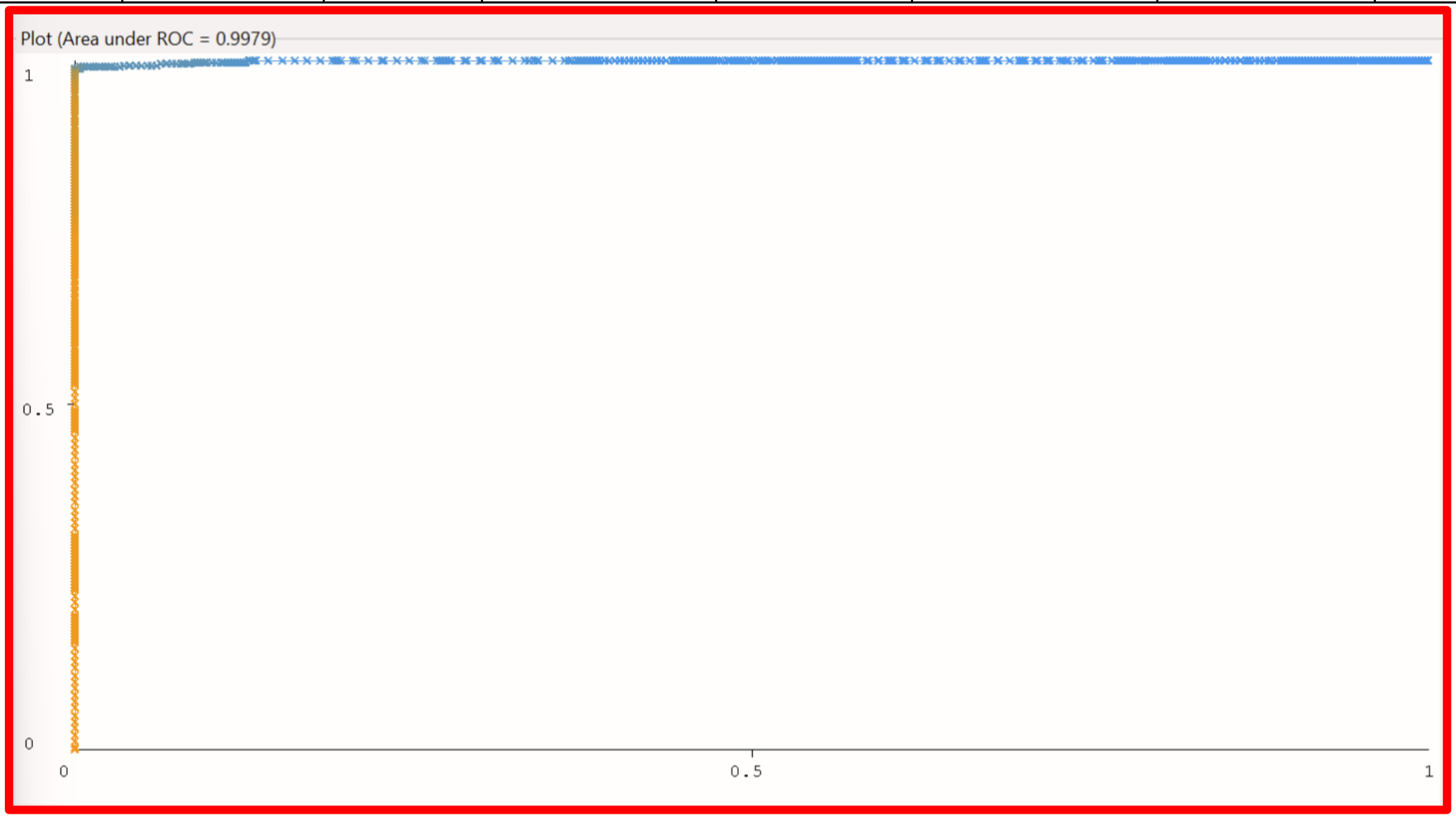

**Figures 21** and **22** *Twenty-fold LogitBoost classifier detailed accuracy report per exploit*

| TP Rate | FP Rate | Precision | Recall | F-Measure | MCC | ROC Area | PRC Area | Class |
|---|---|---|---|---|---|---|---|---|
| 0.985 | 0.001 | 0.999 | 0.985 | 0.992 | 0.983 | 0.998 | 0.999 | DDoS |
| 0.999 | 0.015 | 0.982 | 0.999 | 0.991 | 0.983 | 0.999 | 0.998 | DoS |
| 0.992 | 0 | 0.996 | 0.992 | 0.994 | 0.994 | 1.000 | 0.998 | Reconnaissance |
| 0.514 | 0 | 0.833 | 0.514 | 0.635 | 0.654 | 0.993 | 0.698 | Normal |
| 0.846 | 0 | 1.000 | 0.846 | 0.917 | 0.920 | 1.000 | 0.947 | Theft |

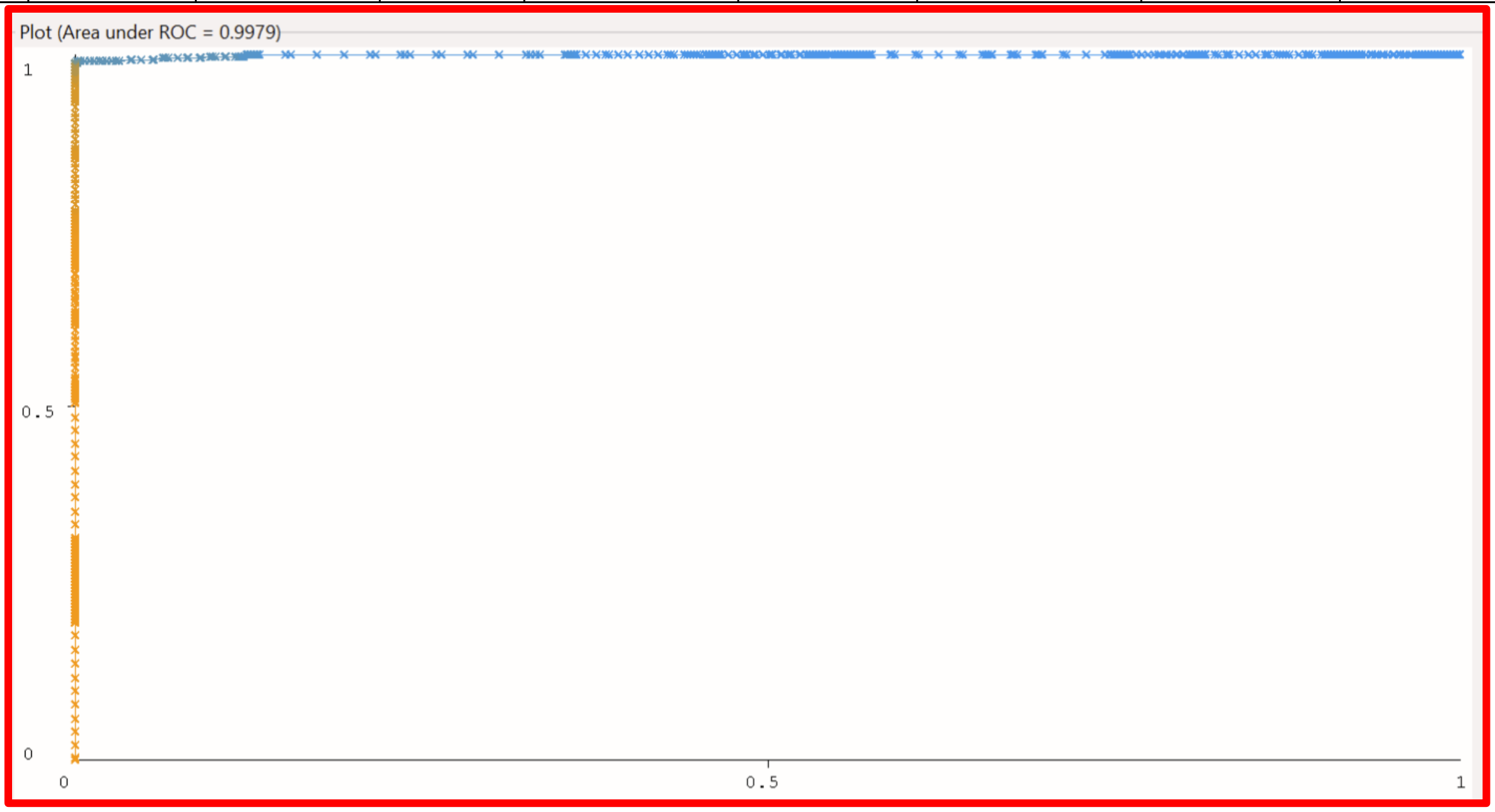

**Figures 23** and **24** *Ten-fold LogitBoost classifier detailed accuracy report per exploit.*

| TP Rate | FP Rate | Precision | Recall | F-Measure | MCC | ROC Area | PRC Area | Class |
|---|---|---|---|---|---|---|---|---|
| 0.985 | 0.001 | 0.999 | 0.985 | 0.992 | 0.983 | 0.998 | 0.999 | DDoS |
| 0.999 | 0.015 | 0.982 | 0.999 | 0.991 | 0.983 | 0.999 | 0.998 | DoS |
| 0.992 | 0 | 0.996 | 0.992 | 0.994 | 0.994 | 1.000 | 0.999 | Reconnaissance |
| 0.514 | 0 | 0.848 | 0.514 | 0.640 | 0.660 | 0.995 | 0.686 | Normal |
| 0.831 | 0 | 1.000 | 0.831 | 0.908 | 0.911 | 1.000 | 0.941 | Theft |

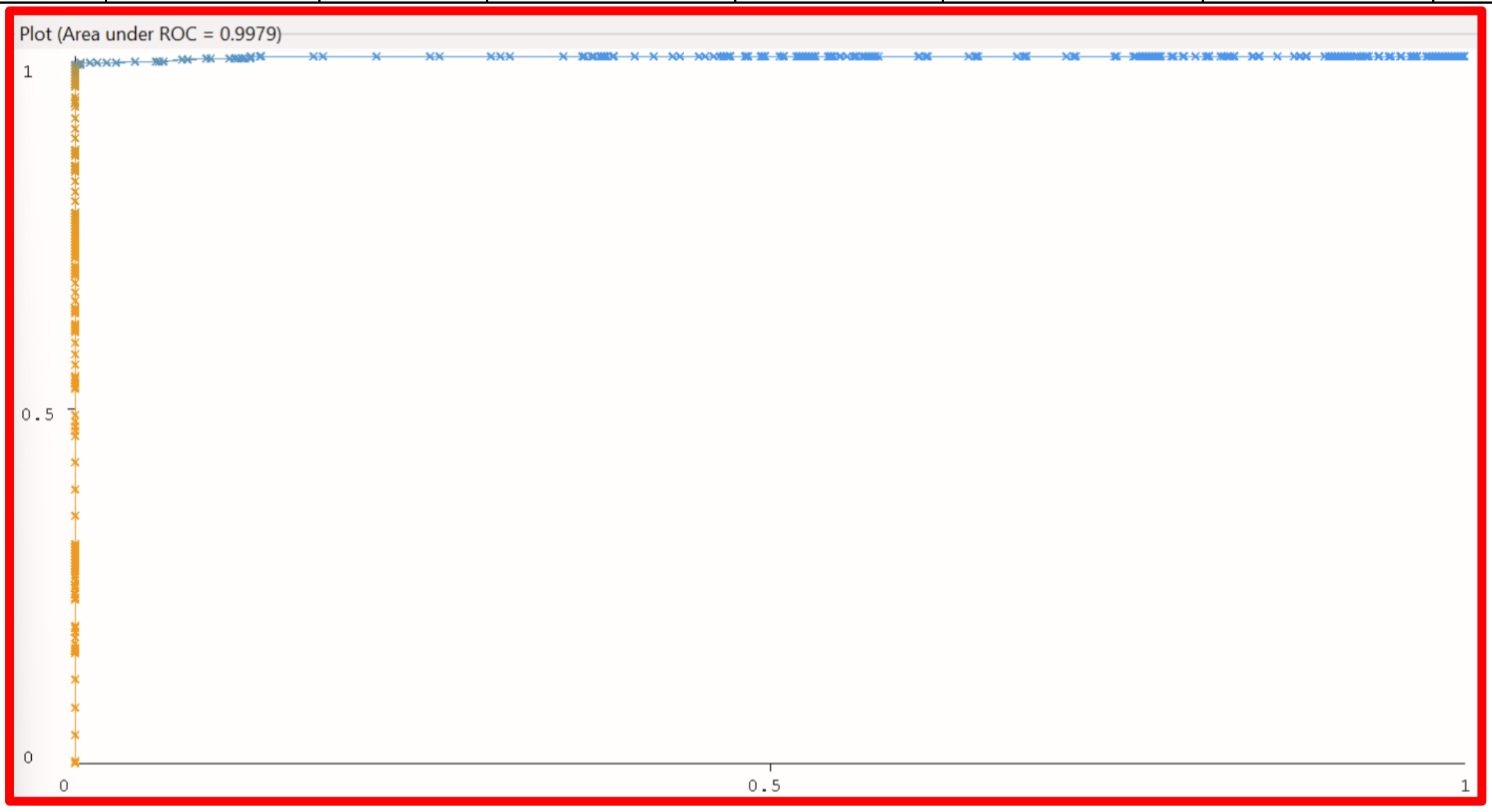

**Figures 25** and **26** *Five-fold LogitBoost classifier detailed accuracy report per exploit.*

Figures 9-26 provide accurate details of Naïve Bayes, UltraBoost, and LogitBoost classifiers and compare themselves on a twenty-fold, ten-fold, and five-fold, respectively. Receiver Operating Characteristic (ROC) graphs are provided in the DDoS class. ROC graphs show the performance of a certain classification model. For the graph, there are two parameters that justify the ROC curve, which is the True Positive (TP) Rate, and the False Positive Rate (FP). The TP rate, synonymous with the term "recall" is defined as the rate at which the classifier correctly identified an instance belonging to that value. The TP rate is calculated by the number of True Positive (TP) instances, divided by the number of True Positives plus the False Negatives [41,43]. The FP rate is defined as the portion of incorrect predictions that were made from the classifier, out of the entire dataset [41,42]. Precision is defined as the portion of the total positive estimates, that was accurate [40] . The F-Measure is the harmonic mean between the precision and the recall [44]. The Matthew correlation coefficient (MCC) is used to reflect the accuracy, including the deficiency of classifiers in a dataset [47]. The ROC Area determines the performance of the classifier under all possible thresholds [45]. The PRC is like the ROC area, except that it is a better measure for datasets that are imbalanced [46]. Each visual figure represents the ROC curve for DDoS for its respective classifier. ROC curves work via thresholds. Thresholds are considered limits that determine whether an instance belongs in a positive class or not. Many graphs may appear different, yet still have a similar AUC (Area Under Curve) because the appearance of the graph is made via different threshold points. As we can see for the DDoS value, Log IT boost has the ideal ROC curve, with an AUC of 0.9979, compared to Naïve Bayes with 0.9577 AUC under 10-folds, and UltraBoost with 0.9717 under ten-folds.

## VI. Conclusion

When conducting our experiments, we tested all algorithms with multiple iterations of folds to identify any growing pattern. Initially, we started with Naïve Bayes with 10-fold cross-validation. The results of the correct prediction came out to be 93.04%. I hypothesized that changing the number of folds in both directions would either increase or decrease the percentage by at least a couple of integers. When performing the same analysis, but with 20 folds as opposed to 10, the accuracy would change with a 0.1% difference. The same outcome applies to Naïve Bayes with 5-fold cross-validation. The building of models would take about one minute per model, no matter the number of folds. When testing with the UltraBoost classifier, I hypothesized that the number of folds would not impact accuracy. However, there was no assumption of what the potential accuracy may be. When performing the test, the time to create the models was exponentially faster than other algorithms used. The time of creation took 296.44 seconds (about 5 minutes). Additionally, the accuracy was sacrificed by 5%, bringing the accuracy to 88.44% on 10 folds, 88.02% on 5 folds, and 88.26% on 20 folds. For the Logitboost classifier, the time to build the model was slightly shorter compared to UltraBoost, but still considerably longer than Naïve Bayes, yet accomplishing its task as being the most accurate classifier out of the three. With an accuracy of 99.123% on the 10-fold test, 99.12136% on the 5-fold test, and 99.121% on the 20-fold test. LogitBoost is the most accurate classifier out of the three used. With Naïve Bayes, being the 2$^{nd}$ most accurate, the potential for Machine Learning to make an impact not only on 5G but on all networks is high. Machine learning will be key for all cybersecurity technologies and needs, and its contribution to 5G should not go unnoticed, especially when IoT devices become more popular in 5G networks.